\documentclass{PoS}
\usepackage{epsfig}

\PoS{PoS(LAT2005)103}

  \newcommand{\bq}{\begin{equation}}
  \newcommand{\eq}{\end{equation}}
  \newcommand{\bqa}{\begin{eqnarray}}
  \newcommand{\eqa}{\end{eqnarray}}

  \newcommand{\nn}{\nonumber}

  \newcommand{\ga}{\gamma}
  
  \newcommand{\ep}{\epsilon}

  \newcommand{\tr}{{\rm tr~}}

\title{A (P)HMC algorithm for $N_{f} = 2 + 1 + 1$ flavours
  of twisted mass fermions}

\ShortTitle{A $N_{f} = 2 + 1 + 1$~ (P)HMC algorithm } 

\author{\speaker{Thomas Chiarappa}\thanks{Supported by the German DFG
        (Deutsche Forschungsgemeinschaft).}\\ 
        Universit\`{a} Milano Bicocca and INFN Milano\\
        E-mail: \email{Thomas.Chiarappa@mib.infn.it}}
\author{Roberto Frezzotti\thanks{From September 1st, 2005,
        new affiliation: Universit\`a di Roma ``Tor
        Vergata'' and INFN Roma2; new e-mail address: Roberto.Frezzotti@roma2.infn.it.} \\ 
        Universit\`{a} Milano Bicocca and INFN Milano\\
        E-mail: \email{Roberto.Frezzotti@mib.infn.it}}
\author{Carsten Urbach\\
        Freie Universit\"{a}t Berlin\\
        E-mail: \email{Carsten.Urbach@physik.fu-berlin.de}}

\abstract{We present a detailed design of a (P)HMC simulation algorithm for
$N_f = 2 + 1 + 1$ maximally twisted Wilson quark flavours. The algorithm
retains even/odd and mass-shift preconditionings combined with multiple
Molecular Dynamics time scales for both the light mass degenerate, $u$
and $d$, quarks and the heavy mass non-degenerate, $s$ and $c$, quarks. 
Various non-standard aspects of the algorithm are discussed, among which those
connected to the use of a polynomial approximation for the inverse (square
root of the squared) Dirac matrix in the $s$ and $c$ quark sector.}

\FullConference{XXIIIrd International Symposium on Lattice Field Theory\\
		 25-30 July 2005\\
		 Trinity College, Dublin, Ireland}

\begin{document}

\section{Introduction and motivation}

A realistic setup for studying many non-perturbative QCD properties 
is obtained by introducing two 
quark pairs, a light ($l$) mass degenerate one ($u$ and $d$ flavours: $2$) and 
a heavier ($h$) mass non-degenerate one ($s$ and $c$ flavours: $1+1$):
for short $N_{f} = 2 + 1 + 1$.
Monte Carlo simulations in this setup are becoming
increasingly feasible in the tmLQCD formulation~\cite{Wtm1,
Wtm2,REV_TM} with action
\bq
S = S_{G}[U] + \bar{\psi}_{l} ~D_{l} [U]~ \psi_{l} + \bar{\psi}_{h}
~D_{h} [U]~ \psi_{h} ~~~~ , ~~~~  \psi_{l} = [u , d] ~~~~ , ~~~~
\psi_{h} = [s , c] \; . \label{LATT_ACT}
\eq
Here $S_{G}$ is a suitable pure gauge action~\footnote{In unquenched 
computations the choice of $S_g[U]$ is important for the phase 
structure of the lattice model~\cite{REV_TM,4TALKUNIF,MONTV_DBW2}.}, 
the Dirac operators read (in the ``physical'' quark basis) 
\bqa
D_{l} = \ga
\tilde{\nabla} - i \ga_{5} \tau_{1} W_{cr} + \mu_{l} \; , &  
\qquad D_{h} = \ga \tilde{\nabla} - i \ga_{5} \tau_{1} W_{cr} + \mu_{h} -
 \ep_{h} \tau_{3} \; , \label{Dirac_op}  \\ 
W_{cr} = -\frac{a}{2} \nabla^{*} \nabla + M_{cr} \; , & 
M_{cr} = \frac{1}{2 \kappa_{cr}} - 4 \; , \label{W_crit}
\eqa
and $\ep_{h}$ is the bare $s$--$c$ quark mass splitting. Among 
the general features of the above formulation~\cite{REV_TM} we
recall: automatic O($a$) improvement~\cite{Wtm2},
robust quark mass protection against ``exceptional
configurations''~\cite{Wtm1}, expected moderate CPU-cost
for unquenched simulations (assuming metastability
problems related to the lattice phase structure are solved)~\cite{REV_TM}. 
The determinant of $D_{h}$ (eq.~(\ref{Dirac_op})) is
real and positive provided $|\ep_{h}| < |\mu_{h}|$, 
which, if $Z_P/Z_S < 1$, induces some limitation
on the renormalised quark
masses, $\hat{m}_{c,s} = Z_{P}^{-1} (\mu_{h} \pm
\frac{Z_{P}}{Z_{S}} \ep_{h})$, one can simulate. 

Through rescaling and chiral rotations of the quark fields (which
do not affect the fermion determinant, besides an irrelevant constant
factor) and by setting $\tilde{\mu}_{l,h} = 2 \kappa_{cr} \mu_{l,h}
~,~ \tilde{\ep}_{l,h} = 2 \kappa_{cr} \ep_{l,h}$, the lattice Dirac 
operators (eq.~(\ref{Dirac_op})) can be rewritten in a form more convenient 
for MC simulations:
\bq
(\tilde{D}_l)_{2 \times
  2} = \left[ \ga \tilde{\nabla} + W_{cr} \right] 2
\kappa_{cr} + i \tilde{\mu}_l \ga_{5} \tau_{3}
\; , \qquad
(\tilde{D}_h)_{2 \times
  2} = \left[ \ga \tilde{\nabla} + W_{cr} \right] 2
\kappa_{cr} + i \tilde{\mu}_h \ga_{5} \tau_{3} + \tilde{\ep}_h \tau_{1}
\; .
\eq
As the value of $\kappa_{cr}$ is {\it a priori} unknown,
$\kappa_{cr}$ is generically replaced by $\kappa$ 
in the definition
of $\tilde{\mu}_{l,h}$ and $\tilde{\ep}_{l,h}$~\footnote{Then
tmLQCD at maximal twist is obtained by setting $\kappa$ to a sensible
estimate of $\kappa_{cr}$ for all quark pairs~\cite{REV_TM}.}. 
With these premises, we focus on the $\ga_{5}$-Hermitian
partner of the Dirac operator(s) above, 
\bqa
\tilde{Q}_{2 \times 2} = \left[
\begin{array}{cc}
\tilde{Q} + i \tilde{\mu} & \tilde{\ep} \ga_{5} \\
\tilde{\ep} \ga_{5} & \tilde{Q} - i \tilde{\mu}
\end{array}
\right] \equiv \left[
\begin{array}{cc}
\tilde{Q}_{+} & \tilde{\ep} \ga_{5} \\
\tilde{\ep} \ga_{5} & \tilde{Q}_{-}
\end{array}
\right] \; , \qquad \tilde{Q} \equiv \ga_{5} \left( \ga
\tilde{\nabla} -\frac{a}{2} \nabla^{*} \nabla + 
\frac{1}{2 \kappa} - 4 \right) 2 \kappa \; ,
\label{Q-def}
\eqa
where we have for the moment dropped the quark pair labels $l$ and $h$.
In the mass degenerate case ($\tilde{\ep}=0$),
a standard HMC algorithm can be employed in view of 
\bqa
\det[\tilde{Q}_{2 \times 2}] = \det[\tilde{Q}_{+}\tilde{Q}_{-}] =
\det[\tilde{Q}_{cr}^{2} + \mu^{2}] ~ \Leftrightarrow ~ \int
\mathcal{D}\phi~ e^{-\phi^{\dag} (\tilde{Q}_{+}\tilde{Q}_{-})^{-1}
  \phi} \; , \nn
\eqa 
with $\phi$ a single flavour pseudofermion field. This fits well
our needs for the $l$ quark pair.
In the mass non-degenerate case ($\tilde{\ep} \neq 0$)
no plain HMC algorithm is straightforwardly applicable because 
\bq
\det[\tilde{Q}_{2 \times 2}] = \det[\tilde{Q}_{+} \ga_{5}
  \tilde{Q}_{-} - \tilde{\ep}^{2} \ga_{5}] =
  \det[\tilde{Q}_{+}\tilde{Q}_{-} - \tilde{\ep}^{2}\tilde{Q}_{+}
  \ga_{5} \tilde{Q}_{+}^{-1} \ga_{5}] \; ,  \label{ND22DET}
\eq
cannot be reproduced by $\int \mathcal{D}\phi~ e^{-\phi^{\dag} (A
A^{\dag})^{-1} \phi}$, with $A$ a 1-flavour matrix. Owing to
the flavour non-diagonal structure of the matrix~(\ref{Q-def}), 
we propose to deal with a two-flavour matrix, $\tilde{Q}_{2 \times 2}
\tilde{Q}_{2 \times 2}^{\dag}$, acting on a two-flavour 
pseudofermion field, $\Phi$, and to use
a polynomial $P(s) \simeq 1/\sqrt{s}$, which gives 
\bq
\det[\tilde{Q}_{2 \times 2}] ~ \Leftrightarrow ~
\int \mathcal{D}\Phi~ e^{-\Phi^{\dag} (\tilde{Q}_{2 \times
    2}\tilde{Q}_{2 \times 2}^{\dag})^{-1/2} \Phi} ~\simeq~
     \int \mathcal{D}\Phi~ e^{-\Phi^{\dag}
         P(\tilde{Q}_{2 \times 2}\tilde{Q}_{2 \times 2}^{\dag}) \Phi}
     \; , \qquad \Phi = \left(
\begin{array}{cc}
\phi' \\
\phi''
\end{array} 
\right) \; ,
\label{first-ND}
\eq 
at least in the $\Phi$-heatbath and the reweighting (or acceptance)
correction. A PHMC algorithm~\cite{PHMC_0,PHMC} appears thus a natural choice 
to include the effects of the $h$ quark pair in MC simulations.

\section{A preconditioned (P)HMC algorithm for $N_f=2+1+1$ flavours}

We outline here the mixed HMC-PHMC (briefly (P)HMC) 
algorithm we are developing, which incorporates even-odd (EO)
and mass-shift preconditionings~\cite{Mass-prec}.
Using different Molecular Dynamics (MD) time steps for different
MD force contributions we expect to obtain a good performance,
in line with that recently achieved in simulations with two Wilson
quark flavours~\cite{Lu0409,UJSW}. 
\\
\vspace*{-0.1cm}

Denoting by $\Pi$ the momenta conjugated to the gauge field, the 
MD-Hamiltonian reads
\bqa
H_{2 + 1 + 1} \quad &=& \quad\frac{1}{2} \Pi \cdot \Pi + S_{G} +
 \Phi^{\dag}_{h}~ P\left( \hat{Q}_{h} 
 \hat{Q}_{h}^\dagger \right)_{h} \Phi_{h} + 
\nn \\ 
\vspace*{0.1cm}
&+& \phi_{1}^{\dag} \left[ \hat{Q}_{l}' 
    (\hat{Q}_{l}')^\dagger  \right]^{-1} \phi_{1} +
 \phi_{2}^{\dag} \left\{ \hat{Q}_{l}' \left[ \hat{Q}_{l}''
 (\hat{Q}_{l}'')^\dagger  \right]^{-1} (\hat{Q}_{l}')^\dagger 
 \right\} \phi_{2} \; ,
\label{MDHa}
\eqa 
where $P( \hat{Q}_{h} \hat{Q}_{h}^\dagger )$ is
a polynomial in $\hat{Q}_{h} \hat{Q}_{h}^\dagger$ approximating
$[\hat{Q}_{h} \hat{Q}_{h}^\dagger]^{-1/2}$ (see
sect.~\ref{DETAIL_POL}) 
while in the $l$ quark sector mass-shift preconditioning 
is applied~\cite{4TALKUNIF,UJSW} on top of EO preconditioning (see
eq.~(\ref{EO_Ql1}) and eq.~(\ref{EO_Ql2})). 
In this way we need two pseudofermions, $\phi_1$ and $\phi_2$,
for the $l$ quark sector and a two-flavour pseudofermion field,
$\Phi_{h}$, for the $h$ quark sector. The EO preconditioned Dirac
operators are 
\vspace*{0.2cm}
\bqa
\hat{Q}_{h} = \ga_{5} \left[
\begin{array}{cc}
1 + i\tilde{\mu}_h\ga_{5} - \frac{M_{oe} (1 - i\tilde{\mu}_h\ga_{5})
    M_{eo}}{1 + \tilde{\mu}_h^{2} - \tilde{\ep}_h^{2}} &  \tilde{\ep}_h
    \left( 1 + \frac{M_{oe} M_{eo}}{1 + \tilde{\mu}_h^{2} -
    \tilde{\ep}_h^{2}} \right) 
\\
 \tilde{\ep}_h \left( 1 + \frac{M_{oe} M_{eo}}{1 + \tilde{\mu}_h^{2} -
    \tilde{\ep}_h^{2}} \right) & 1 - i\tilde{\mu}_h\ga_{5} - \frac{M_{oe}
    (1 + i\tilde{\mu}_h\ga_{5}) M_{eo}}{1 + \tilde{\mu}_h^{2} -
    \tilde{\ep}_h^{2}} 
\end{array}
\right] \; ,
\label{EO_Q22}
\eqa 
\bqa
(M_{eo(oe)})_{x,y} = -\kappa \sum_{\mu} \left[ (1 + \ga_{\mu})
  U^{\dag}_{\mu}(y) \delta_{y, x - \hat{\mu}} + (1 - \ga_{\mu})
  U_{\mu}(x) \delta_{y, x + \hat{\mu}} \right]  \; ,
\label{EO_hopping}
\eqa 
with $\hat{Q}_h$ having a $2 \times 2$ flavour structure that is made 
apparent in the r.h.s. of eq.~(\ref{EO_Q22}), and 
\bqa
\hat{Q}_{l}' & = & \ga_{5} \left[ 1 + i \left( \tilde{\mu}_l +
  \delta\tilde{\mu}_l \right) \ga_{5} - \frac{M_{oe} (1 - i\left(
\tilde{\mu}_l +
  \delta\tilde{\mu}_l \right)\ga_{5})
    M_{eo}}{1 + \left( \tilde{\mu}_l +
  \delta\tilde{\mu}_l \right)^{2}} \right] \; ,
\label{EO_Ql1} 
\eqa 
\bqa
\hat{Q}_{l}''& = & \ga_{5} \left[ 1 + i \tilde{\mu}_l \ga_{5} - \frac{M_{oe}
    (1 - i \tilde{\mu}_l \ga_{5}) M_{eo}}{1 + \tilde{\mu}_l^{2}} \right]
\label{EO_Ql2} \; ,
\eqa
with $\hat{Q}_{l}'$ and $\hat{Q}_{l}''$ carrying the shifted 
($\tilde{\mu}_l + \delta \tilde{\mu}_l$) and the physical ($\tilde{\mu}_l$) 
twisted mass parameters. We remark that (due to the absence of the 
Sheikholeslami--Wohlert term in the fermionic action)
the gauge field enters the Dirac matrices eqs.~(\ref{EO_Q22}),
(\ref{EO_Ql1}) and (\ref{EO_Ql2}) only through $M_{eo}$ and $M_{oe}$, 
eq.~(\ref{EO_hopping}). It follows that the evaluation of the MD driving 
force $\dot{\Pi} = - \delta_{U} H_{2 + 1 + 1}  $
can be, as usual, traced back to that of $\delta_{U} M_{eo}$
and $\delta_{U} M_{oe}$ plus the (many) necessary applications of
the relevant Dirac matrices.
\\

With the (P)HMC update~\cite{PHMC,UJSW} dictated by the
Hamiltonian $H_{2 + 1 + 1}$ (eq.~(\ref{MDHa})) one ends (after
thermalisation) with a sample of gauge configurations equilibrated
with respect to the effective gauge action 
$S_G[U] - \log |\det \hat{Q}''_l|^2  +
\log \det P( \hat{Q}_h \hat{Q}_h^\dagger )$. 
A way to correct for the polynomial approximation of 
$[ \hat{Q}_h \hat{Q}_h^\dagger ]^{-1/2}$ is to reweight 
all the observables
${\cal O}={\cal O}[U]$ with a correction factor~\cite{PHMC} 
that provides a noisy
estimate of ~$\det [ \hat{Q}_h \hat{Q}_h^\dagger ]^{1/2}
P (\hat{Q}_h \hat{Q}_h^\dagger) =
\det \hat{Q}_h P (\hat{Q}_h \hat{Q}_h^\dagger) $ 
(see also sect.~\ref{DETAIL_POL}).

\subsection{MD force contributions and multiple time scales} \label{3B}

The MD driving force can be defined (omitting for brevity all indices)
as 
\bq
\dot{\Pi} = - \delta_{U} H_{2 + 1 + 1} \equiv \tr \left[ \delta U
  F + F^{\dag} \delta U^{\dag}
  \right] \; , \qquad F = F_G + F_h + F_{l1} + F_{l2} \; ,
\eq
where (see eq.~(\ref{MDHa})) the pure gauge contribution $(F_{G})$ 
is completely standard, the $l$ quark sector contributions ($F_{l1}$
and $F_{l2}$) can be straightforwardly evaluated following 
Refs.~\cite{UJSW,Alg+MetaSt} and also the $h$ quark sector contribution 
($F_h$) poses no principle problems (see sect.~\ref{DETAIL_POL} for
more details). 

Based on Refs.~\cite{Lu0409,UJSW}, we expect that for typical choices 
of $S_G$, the lattice spacing and the quark masses, one can have
a hierarchy in the average ($av$) size of the individual force
contributions, 
\bq
|F_{G}|_{av} ~ > ~ |F_{l1}|_{av} ~ > ~ |F_{l2}|_{av} \; ,
\qquad  |F_{l1}|_{av} ~ \geq ~ |F_{h}|_{av} ~ \geq ~ |F_{l2}|_{av} \; ,
\label{F_HIER}
\eq
provided $\tilde{\mu}_l >0$ is not too small and
$\delta \tilde{\mu}_l>0$ is appropriately chosen~\cite{UJSW}. 
$|F_{h}|_{av}$ is expected to be small for heavy
quark masses, i.e.\ large $|\tilde{\mu}_h|$ and
$|\tilde{\mu}_h| > |\tilde{\ep}_h|$~\footnote{
It is important that $|\tilde{\ep}_h|$ is not too close
to $|\tilde{\mu}_h|$. In fact, for 
$|\tilde{\ep}_h| \geq |\tilde{\mu}_h|$ not only the
positivity of the determinant is no longer guaranteed, but
the matrix $\hat{Q}_h$ (as
well as $D_h$) can even develop zero eigenvalues.}. 

The hierarchy in eq.~(\ref{F_HIER}) --once realised-- suggests that
an optimal performance is obtained by implementing a MD leapfrog scheme
where the force contributions $(F_{G},F_{h},F_{l1},F_{l2})$ enter 
associated to different time steps $(\delta\tau_{G},\delta\tau_{h},
\delta\tau_{l1},\delta\tau_{l2})$ so as to get  
\bq
|F_{G}|_{av} \delta \tau_{G} \simeq
|F_{h}|_{av} \delta \tau_{h} \simeq 
|F_{l1}|_{av} \delta \tau_{l1} \simeq 
|F_{l2}|_{av} \delta \tau_{l2} \; ,
\qquad \delta \tau_{i} N_{i} \equiv \tau_{traj} \; , \quad i=\{G,l1,l2,h\} \; ,
\label{TAUSTEPS}
\eq
with $(N_{G},N_{h},N_{l1},N_{l2})$ a set of integers and $\tau_{traj} \sim 1$
the time length of a MD trajectory.

\subsection{Some possible algorithmic variants}

Several modifications of the above (P)HMC algorithmic scheme are 
of course possible.
For instance, the 
correction for the polynomial approximation
can be moved, fully or partially, from the reweighting to a modified
A/R Metropolis step~\cite{Noisy,TSMB}. 
If this is done ``fully'' the modified A/R step
compensates for both $\det[\sqrt{\hat{Q}_h \hat{Q}_h^\dagger}
P(\hat{Q}_h \hat{Q}_h^\dagger)] \neq 1$ and
the finite MD time step(s).

The update of the gauge field itself can be performed by means
of a Multiboson-like algorithm~\cite{LUSCHER_MB} using suitable polynomials
to approximate the appropriate power of the inverse Hermitean 
Dirac matrices  (see e.g.\ Ref.~\cite{TSMB} for more details)
relevant for the $l$ and $h$ quark sectors. 

Another interesting possibility is to employ
a non-standard HMC algorithm, whose MD is guided by an Hamiltonian
$\tilde{H} \neq H_{2+1+1}$ such that a good acceptance is still
obtained in the A/R test. One might try e.g.\ an $\tilde{H}$ that
differs from $H_{2+1+1}$ (eq.~(\ref{MDHa})) only by the replacement of
$ \Phi^{\dag}_{h}~ P( \hat{Q}_{h}
 \hat{Q}_{h}^\dagger )_{h} \Phi_{h} $
with
$ \Phi^{\dag}_{h}~  \hat{Q}_{h}^{-1} \Phi_{h} $. 
Numerical experience is of course crucial to test these 
possible variants and choose the most efficient one.

\section{Polynomial approximations for the $h$ quark sector 
($s$ and $c$ flavours)} \label{DETAIL_POL}

The well known Chebyshev polynomial approximation method allows 
to approximate the inverse square root of the operator
$\hat{S} = \rho_h \hat{Q}_h \hat{Q}_h^\dagger $,
where $\rho_h$ is a positive normalisation factor
such that on ``practically all'' gauge configurations
the highest eigenvalue of $\hat{S}$, say $s_H$,    
satisfies $0.8 \leq s_H < 1$. The polynomial $P(\hat{S})=
P_{n,s_L}(\hat{S})$ of even degree $n$ in $\hat{S}$ that is designed
to approximate $\hat{S}^{-1/2}$ in the eigenvalue interval
$[s_L, 1]$~\footnote{By $s_L$ we denote the lowest eigenvalue of $\hat{S}$.
It can be estimated in the early stages of a simulation, e.g.\ by starting
with a ``trial'' polynomial $P_{N , \lambda}$ and then changing 
$N$ and $\lambda$ on the basis of on-the-fly measurements of 
eigenvalues of $\hat{S}$.} 
can be written 
(via a product representation with a normalisation
factor ${\cal N} >0$ and roots $z_k$, $k=1,\dots ,n$ such
that $z_{n+1-k} = z_k^*\;$)  
in a manifestly positive form 
\bq
P_{n,s_L}(\hat{S}) = {\cal N} \prod_{k=1}^n (\hat{S} - z_k)
\quad \Rightarrow \quad P_{n,s_L}
\equiv B_{n/2,s_L} B_{n/2,s_L}^\dagger 
= \hat{S}^{-1/2} \left[1 + R_{n,s_L} \right] \; ,
\label{P_FORM}
\eq

\begin{figure}[!htbp]
\begin{center}
\epsfig{file=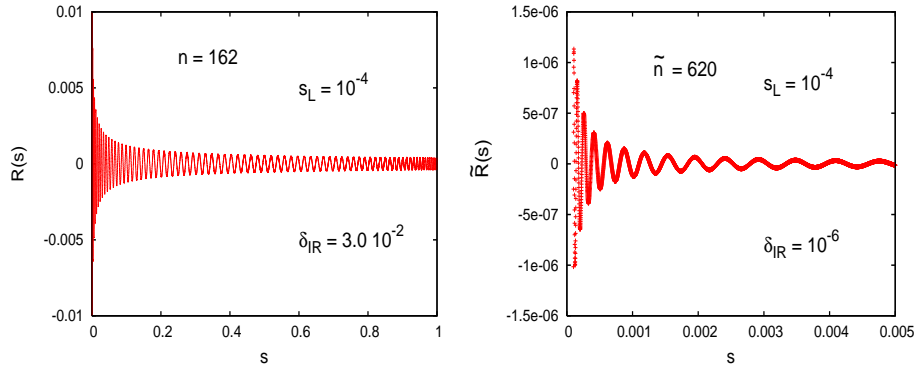,height=12.0cm,width=4.75cm,angle=270}
\vspace*{-0.1cm}
\caption{$R(s)=R_{n,s_L}(s)$ for $(n,s_L)=(162,10^{-4})$ (left)
and $\tilde{R}(s)=\tilde{R}_{\tilde{n},s_L}(s)$ for $(\tilde{n},s_L)
=(620,10^{-4})$ (right).} 
\label{plot1}
\end{center}
\vspace*{-0.6cm}
\end{figure}

\vspace*{-0.1cm}
\hspace*{-0.8cm}
where $R_{n,s_L}= R_{n,s_L}(\hat{S})$ is called the relative fit error. 
Denoting by $s$ a generic eigenvalue of $\hat{S}$, in
Fig.~\ref{plot1} (left panel) we plot for illustration
the relative fit error $R_{n,s_L}(s)$ for $s_L=10^{-4}$
and a value of $n$, namely $n=162$, taken such that 
$|R_{n,s_L}(s)| \leq 0.03$. The chosen value of $s_L$ is rather
conservative, since it is very low as compared to those that, based on
simulation experience with two mass degenerate quarks, we expect to
have to face in the $h$ quark sector while working with realistic
parameters. We see from Fig.~\ref{plot1} that $R_{n,s_L}(s)$ tends to 
increase when decreasing $s \in [s_{L},1]$, which we believe is acceptable 
in view of the expected non-high 
density of eigenvalues in the low end of the spectrum of $\hat{S}$.
A conservative and an ``effective'' measure of the magnitude of
the relative fit error are thus given, respectively, by
$\delta_{IR} \equiv \max_{s\in[s_{L},1]} |R_{n,s_L}(s)| = |R_{n,s_L}(s_L)|$ and
$\delta_{UV} \equiv \max_{s\in[0.5,1]} |R_{n,s_L}(s)|$
with $\delta_{UV}$ smaller than $\delta_{IR}$ (typically by 
an order of magnitude). 

\vskip -0.2cm
\hspace*{-0.8cm}

For the task of computing the contribution $F_h$ to the MD driving force we
plan to exploit the product representation of $P_{n,s_L}(\hat{S})$, see 
eq.~(\ref{P_FORM}), and then proceed analogously to Ref.~\cite{PHMC}.  
A careful ordering~\cite{Ordering} of the $n$ monomials 
in $\hat{S}$ together with 64 bit precision should be sufficient to keep
rounding errors under control for polynomials with $n$ up to one thousand.
\\
\vspace*{-0.1cm}

There are two places where a second polynomial approximation (or an
equivalent method such as a rational CG solver~\cite{Bunk}) is needed.
The first place is the generation of $\Phi_h$ distributed according to 
$\exp [-\Phi_h^\dagger P_{n,s_L}(\hat{S}) \Phi_h]$, see eq.~(\ref{MDHa}).
If $r_h$ is a random Gaussian (two flavour) vector, we can construct 
$\Phi_h = B_{n/2,s_L}^{-1}(\hat{S}) ~r_h~$ by e.g.\ evaluating 
\vspace{-0.1cm}
\bq
\Phi_h = \tilde{P}_{\tilde{n},s_L}(\hat{S}) 
B_{n/2,s_L}^\dagger(\hat{S}) ~[ \sqrt{\rho_h} \hat{Q_h} ]
~r_h \; , \qquad \tilde{P}_{\tilde{n},s_L}(\hat{S}) =
\left[ \sqrt{\hat{S}} P_{n,s_L}(\hat{S}) \right]^{-1}
\left[ 1 + \tilde{R}_{\tilde{n},s_L}(\hat{S}) \right] \; ,
\label{BHB_PTIL}
\eq

\vspace*{-0.1cm}
\hspace*{-0.8cm}
where $\tilde{P}_{\tilde{n},s_L}(\hat{S})$ is a new polynomial 
of degree $\tilde{n}$ in
$\hat{S}$ that approximates $[\sqrt{\hat{S}} P_{n,s_L}(\hat{S})]^{-1}$
with very high precision in the eigenvalue interval $[s_L,1]$. Values
of the corresponding relative fit error, $\tilde{R}_{\tilde{n},s_L}$,
not exceeding in modulus $\tilde{\delta}_{IR} = 10^{-6}$ can be reached for
a degree $\tilde{n} = 620$, in the above mentioned case of
$s_L=10^{-4}$ and $n=162$, as it is illustrated in Fig.~\ref{plot1}
(right panel). The evaluation of $\tilde{P}_{n,s_L}(\hat{S})$ times a
vector can be safely performed in 64 bit arithmetics by using a
recursive relation (such as Clenshaw's) that is stable against
roundoff. 

The other place where the high precision polynomial
$\tilde{P}_{n,s_L}(\hat{S})$ in eq.~(\ref{BHB_PTIL}) 
may be useful is in the evaluation of $W[\eta; U] = \exp\left\{
\eta^{\dag} \left( 1 - [ P_{n,s_L}(\hat{S}) \sqrt{\hat{S}} ]^{-1}
\right) \eta \right\}$, the noisy reweighting factor that is needed to
get the ``exact'' v.e.v. of a generic observable ${\cal O} = {\cal
  O}[U]$ (here $\eta$ is the noise field)~\cite{PHMC} 
\bqa
& \langle {\cal O} \rangle = \int D\mu[U] \int D\eta
~e^{-\eta^\dagger\eta} ~W[\eta;U] ~{\cal O}[U] \;/\; \int D\mu[U] \int
D\eta ~e^{-\eta^\dagger\eta} ~W[\eta;U] \; , 
\nn
\eqa
\bqa
& \quad \int D\mu[U] \equiv \int DU e^{-S_G[U]} 
~\det [ \hat{Q}_{l}'' \hat{Q}_{l}''^{\dag} [U] ] ~\det\left[
  P(\hat{S}[U]) \right]^{-1} ~\equiv~ Z_{(P)HMC} \label{vev-rew} \quad
.
\eqa

\section{Conclusions and Acknowledgements}

We discussed an exact algorithm for $N_f=2+1+1$ flavours of maximally
twisted quarks. Implementation of the algorithm and investigation 
of important numerical properties, such as the spectrum of 
$\hat{Q}_h \hat{Q}_h^\dagger$ and the magnitude of the contribution
$F_h$ to the MD driving force, are in progress. 

We thank K.~Jansen for many valuable discussions and advises.
The work of T. C. is supported by the Deutsche Forschungsgemeinschaft in
the form of a Forschungsstipendium CH398/1.



\end{document}